\documentstyle{aa}

\input{epsf}

\begin{document}
\thesaurus{13.09.1;11.05.2;11.19.3} 

\title{Deep far infrared ISOPHOT survey in "Selected Area 57"}

\subtitle{I. Observations and source counts}

\author{M.J.D. Linden-V{\o}rnle \inst{1,2}
\and H.U. N{\o}rgaard-Nielsen \inst{2}
\and H.E. J{\o}rgensen \inst{1} 
\and L. Hansen \inst{1}
\and M. Haas \inst{3}
\and U. Klaas \inst{3}
\and P. \'{A}brah\'{a}m \inst{3}
\and D. Lemke \inst{3}
\and I. Lundgaard Rasmussen \inst{2}
\and H.W. Schnopper \inst{4}}

\offprints{M.J.D. Linden-V{\o}rnle (michael@astro.ku.dk)}

\institute{Niels Bohr Institute for Astronomy, Physics and Geophysics,
Astronomical Observatory, Juliane Maries Vej 30, DK--2100 K{\o}benhavn {\O},
Denmark
\and Danish Space Research Institute,
Juliane Maries Vej 30, DK--2100 K{\o}benhavn {\O}, Denmark
\and Max-Planck-Institut f{\"u}r Astronomie (MPIA), K{\"o}ningstuhl 17,
D--69117 Heidelberg, Germany
\and Harvard-Smithsonian Center for Astrophysics, 60 Garden Street, 
Cambridge, MA 02138, USA}

\date{Received date; accepted date}

\maketitle
\markboth{M.J.D. Linden-V{\o}rnle et al.: Deep ISOPHOT survey in "Selected Area 57"}{M.J.D. Linden-V{\o}rnle et al.: Deep ISOPHOT survey in "Selected Area 57"}
 
\begin{abstract}
We present here the results of a deep survey in a 0.4 deg$^2$ 
blank field in Selected Area 57 conducted with the ISOPHOT instrument 
aboard ESAs Infrared Space Observatory (ISO\footnote{Based on observations 
with ISO, an ESA project with instruments funded by ESA member states (especially
the PI countries: France, Germany, the Netherlands, and the United Kingdom)
and with the participation of ISAS and  NASA.}) at both 60 $\mu$m and 
90 $\mu$m. The resulting sky maps have a spatial resolution of 15 $\times$ 23
arcsrc$^2$ per pixel which is much higher than the 90 $\times$ 90
arcsec$^2$ pixels of the IRAS All Sky Survey. 
We describe the main instrumental effects encountered in our
data, outline our data reduction and analysis scheme and present astrometry
and photometry of the detected point sources. With a formal
signal to noise ratio of 6.75 we have source detection limits of 90 mJy at
60 $\mu$m and 50 mJy at 90 $\mu$m. To these limits we find cumulated 
number densities of 5$\pm 3.5$ deg$^{-2}$ at 60 $\mu$m and 14.8$\pm 5.0$
deg$^{-2}$ at 90 $\mu$m. 
These number densities of sources are found to be lower than
previously reported results from ISO but the data do not allow us to
discriminate between no-evolution scenarios and various evolutionary
models. 

\end{abstract}

\section{Introduction}
It is widely accepted that a significant part of the evolution of
galaxies is hidden from UV/optical studies due to internal absorption 
by dust grains, and that the absorbed radiation is re-emitted in the infrared. 
  
The IRAS All Sky Survey has revealed more than 25.000 galaxies, of which 
only half were already known at optical wavelengths (Soifer et al.
\cite{soif1}). The vast majority of these are local late-type spirals while
ellipticals and S0 galaxies were rarely detected. Only a few very luminous 
infrared galaxies were detected at significant redshifts. 

A first indication of a possible evolution of the population of infrared 
sources at the IRAS detection limit was found by Hacking \& Houck 
(\cite{hack1}) (HH87 hereafter). They exploited all the data obtained during 
observations of the IRAS secondary calibration source, NGC 6543, a planetary 
nebula close to the North Ecliptic Pole. Although this area 
(6.25 deg$^2$) is far from being the most 'cirrus clean' area in the 
sky, the extensive coverage by IRAS resulted in detection of sources 
(S/N $>$ 5) down to about 50 mJy in the 60 $\mu$m band, about 10 times fainter 
than the detection limit in the IRAS All Sky Survey.

Following the pioneering work by HH87 several other deep IRAS surveys have been 
published. Gregorich et al. (\cite{greg1}) have analysed $\rm \sim 20 ~deg^{2}$ 
-- socalled filler fields, not including the HH87 field --  with eight or more 
observations and designated with 'moderate cirrus' or 'low cirrus' flags in 
the IRAS Faint Source Survey (FSS) (Moshir et al. \cite{mosh1}).  
Gregorich et al. (\cite{greg1}) find the source density at 
$f_{\nu}(60\mu\mbox{m}) = 50$ mJy to be about twice as high as that found by 
HH87. 

Bertin et al. (\cite{ber1}) have extracted 60 $\mu$m data from the FSS from 
$\rm \sim 400 ~deg^{2}$ in four separate contiguous areas selected with good 
coverage, low cirrus indicators and a minimum of nearby galaxies. They have 
performed a detailed analysis of the statistical errors in the number counts 
at the faint limit -- the Eddington bias -- and give corrected number counts 
down to $f_{\nu}(60\mu\mbox{m}) = 100$ mJy. At this limit they find a source 
density about 25\% lower than HH87. They suggest that the large discrepancy 
between their results and Gregorich et al. (\cite{greg1}) is caused by cirrus 
contamination in their fields. 

Analysing the HH87 data, Hacking et al. (\cite{hack2}) found a significant 
excess of their 60 $\mu$m number density below 100 mJy, if no 
evolution out to $z$ $\approx$ 0.2 is assumed. They fit the number densities 
with simple models assuming a power law in $(1 + z)$ either as pure density 
or pure luminosity evolution, but are not able to distinguish between these 
two types of models.     

This excess of sources at faint infrared fluxes have invoked a considerable 
effort in constructing theoretical models for galaxy evolution. 
Guiderdoni et al. (\cite{guid1}) pre\-sent semi-analytic models of galaxy
evolution in the infrared following the non-dissipative and dissipative 
collapses of primordial pertubations, incorporating star formation, stellar 
evolution and feedback as well as absorption of starlight by dust, and its 
re-emission in the infrared. In order to fit the slope of the IRAS number 
densities,  Guiderdoni et al. (\cite{guid1}) introduce a significant population
of luminous dust shrouded galaxies, simulating the ultraluminous infrared 
galaxies (ULIRGs) discovered by IRAS. Still, at 
$f_{\nu}(60\mu\mbox{m}) \approx 100$ mJy, their best model (E) is 20\% below 
the number counts found by Hacking et al. (\cite{hack2}). 
   
Another detailed set of galaxy evolution models in the infrared
range incorporating the energy emitted by various stellar generations 
with different abundances, the opacity of the enriched interstellar gas and
the flux reradiated by dust has been developed by Franceschini et al. 
(\cite{fran}). In order to explain the HH87 60 $\mu$m counts 
Franceschini et al. (\cite{fran}) favour a model including a population of 
early type galaxies heavily obscured in the optical by dust.    

Optical identification of the HH87 sources has been performed by Ashby et al. 
(\cite{ash1}) by associating the sources with the closest object on the Palomar 
Observatory Sky Survey plates  within an error circle of 2 arcmin in diameter.
From optical spectra Ashby et al. (\cite{ash1}) find a strong concentration of 
objects at redshifts around 0.088 and all redshifts being lower than 0.26. The 
redshifts were determined from two or more emission and/or absorption lines, 
but Ashby et al. (\cite{ash1}) do neither provide information about the 
identification of the lines nor the line ratios for a detailed classification 
of the spectra. Only in a few cases there were more than one obvious optical 
counterpart, but they most often have about the same redshift. 

Kawara et al. (\cite{Ka98}) have recently presented results from their
ISOPHOT survey of a 1.1 deg$^2$ area in the Lockman Hole region.
Their survey was conducted at both 90 $\mu$m using the C100 detector
and at 175 $\mu$m using C200. They find a number density of sources
of 33 deg$^{-2}$ at 90 $\mu$m down to 150 mJy and 40 deg$^{-2}$ at 175 
$\mu$m also down to 150 mJy. They find that the number densities at 
175 $\mu$m are 3-10 times higher than the no-evolution model by 
Guiderdoni et al. (\cite{guid1}) depending on details in the calibration.

Very recently, the first results of the FIRBACK far infrared survey with the
ISOPHOT C200 detector array performed at 175 $\mu$m have been
published (Puget et al. \cite{pug1}; Dole et al. \cite{dol1}).  
Dole et al. (\cite{dol1}) find that their 175 $\mu$m
source counts, extracted from a 4 $\rm deg^2$ area  for fluxes down to
about 100 mJy, fits well with the E-model by Guiderdoni et al. (\cite{guid1}), 
while their counts fall in between the Franceschini et al. (\cite{fran2}) 
models with and without evolution.

Also very recently, the initial results of deep surveys at 850 $\mu$m 
with the first submillimeter detector array, SCUBA, on the James Clerk Maxwell
Telescope have been reviewed by Mann (\cite{man1}). The SCUBA 850 $\mu$m source 
counts down to 1 mJy are in line with the E-model by Guiderdoni et al. (\cite{guid1}) 
and a few times higher than the model by Franceschini et al. (\cite{fran2}). 
Due to the limited angular resolution of SCUBA ($\sim$15 arcsec) the optical 
identification is still somewhat dubious, but many of the sources seem to be 
associated with optical objects, which have photometric redshifts in the 
range $1 < z < 3$.

\section{The ISOPHOT observations}
\begin{figure}[ht]
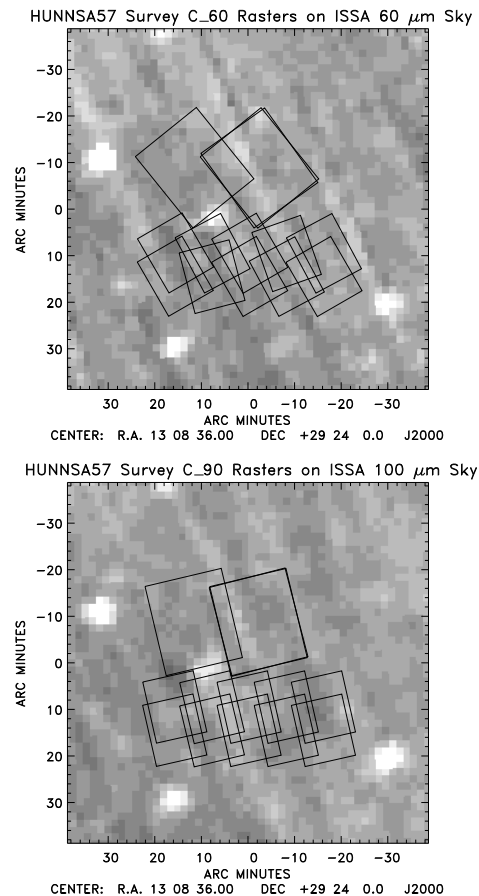

\begin{center}
\leavevmode
\hbox{%
\epsfysize=60mm
\epsfbox{9369.f1a}}
\leavevmode
\hbox{%
\epsfysize=60mm
\epsfbox{9369.f1b}}
\end{center}
\caption{The areas covered by the raster maps of the HUNNSA57 survey for 
the two bands are indicated as overlays on ISSA maps. Note the considerable
overlap between the individual raster maps}
\end{figure}

As part of the Infrared Space Observatory (ISO, Kessler et al. \cite{Ke96}) 
Central Programme, we have performed a deep ISOPHOT survey of a 
$\sim 0.4$ deg$^2$ area within Selected Area 57 (e.g. Stebbins et al. 
\cite{st50}). We have chosen this 
area because 1) it is close to the North Galactic Pole and thus has a low 
far infrared sky background and it is relatively clean with respect to 
cirrus emission, and 2) because Koo et al. (\cite{Koo:Kr}) have identified 
QSO's in this area down to B = 22.5. With the pre-launch estimate of 
the performance of the ISOPHOT instrument it was reasonable to expect, 
that a significant number of these QSO's would be detected in our deep survey. 

As an extension to this work we have also included the deep ISOPHOT survey  
at 90 $\mu$m of a 1.1 deg$^2$ area in the Lockman Hole 
(Kawara et al. \cite{Ka98}).

We have performed the observations using the ISOPHOT C100 detector array, 
both at 60 $\mu$m and at 90 $\mu$m producing three $16 \times 16$ arcmin$^2$
and ten $10 \times 10$ arcmin$^2$ raster maps in each wavelength band. One of 
the large raster maps is observed twice in each band in order to facilitate 
consistency checks. In Fig. 1 the areas covered by the different raster maps 
for the two bands are indicated as overlays on IRAS Sky Survey Atlas (ISSA) 
maps (Beichman et al. \cite{iras}). The total area covered is 0.40 deg$^2$ 
in each band. In order to obtain all data early in the mission, we chose not 
to constrain the orientation of the rasters. As a consequence there are 
considerable overlaps between several of the rasters (see Fig. 1). This 
has given us an additional possibility for investigating the reliability of 
the obtained data. 

The C100 array consists of $3 \times 3$ Ge:Ga detectors, each with an 
angular size on the sky of $43.5 \times 43.5$ arcsec$^2$. Due to gaps 
between the detector pixels only 93\% of the total area of the array is 
covered. These informations, as well as an extensive explanation of how the 
C100 detectors are working and details about observing modes including the 
chopped raster map mode used for the observations described in this work 
(PHT32) can be found in the ISOPHOT Observer's Manual (Klaas et al. \cite{Kl}). 
Further details about the design and performance of the ISOPHOT instrument can
be found in Lemke et al. (\cite{le93}) and (\cite{le94}) whereas the overall 
in-flight performance of ISOPHOT is presented by Lemke et al. (\cite{Le96}). 
We have listed our ISOPHOT observations in Table 1 and the specific instrument 
setup parameters in Table 2.   
\begin{table}
\begin{center}
\begin{tabular}{lcc} \hline
 & \multicolumn{2}{c}{Filter} \\ 
Target       & C\_60    & C\_90    \\ \hline
SA57\_1\_1   & 19800825 & 24300926 \\
SA57\_1\_1   & 20401601 & 24500702 \\
SA57\_1\_2   & 19800903 & 24500804 \\
SA57\_2\_1   & 21401405 & 24601906 \\
SA57\_2\_2   & 21800707 & 24601808 \\
SA57\_2\_3   & 21401309 & 24601710 \\
SA57\_2\_4   & 24300711 & 24500912 \\
SA57\_2\_5   & 21401513 & 24601214 \\
SA57\_2\_6   & 21800815 & 24601516 \\
SA57\_2\_7   & 23402817 & 24601418 \\
SA57\_2\_8   & 21401219 & 24601320 \\
SA57\_2\_9   & 21800921 & 24501022 \\
SA57\_2\_10  & 21401023 & 24601624 \\ \hline
$\alpha$ Aql & 53400412 &    --    \\
HR 1654      &    --    & 65701316 \\ \hline
LH\_NW1      &    --    & 19400201 \\
LH\_NW2      &    --    & 19400202 \\
LH\_NW3      &    --    & 19500103 \\
LH\_NW4      &    --    & 19500104 \\
LH\_E1       &    --    & 20800709 \\
LH\_E2       &    --    & 20800710 \\
LH\_E3       &    --    & 20900111 \\
LH\_E4       &    --    & 20900112 \\ \hline
\end{tabular}
\end{center}
\caption{The ISOPHOT observations used in this work identified by their
Target Dedicated Time (TDT). The first 11 targets are the HUNNSA57 data, 
the two next targets are stars observed for calibration purposes and
the final eight targets are the observations of the Lockman Hole 
(Kawara et al. \cite{Ka98})}
\end{table}
 
\begin{table*}
\begin{center}
\begin{tabular}{ccccccccccccc} \hline
Filter & Mode & AOT & nY & nZ & dY & dZ & NRPR & $t_{read-out}$ & NRCP &
NCHP & CHST & NSWP \\
\hfill & \hfill & \hfill & \hfill & \hfill & [$\arcsec$] & [$\arcsec$] & \hfill & [ms] & \hfill & \hfill & [$\arcsec$] &
\hfill \\ \hline 
C\_60 & Chopping & PHT32 & 11 & 14 & 92  & 69  & 16 & 31.25 & 4 & 13 & 15 & 3 \\  C\_60 & Chopping & PHT32 & 7  & 9  & 92  & 69  & 16 & 31.25 & 4 & 13 & 15 & 4 \\ C\_90 & Chopping & PHT32 & 11 & 14 & 92  & 69  & 8  & 31.25 & 4 & 13 & 15 & 2 \\  C\_90 & Chopping & PHT32 & 7  & 9  & 92  & 69  & 8  & 31.25 & 4 & 13 & 15 & 3 \\
\hline
C\_60 & Chopping & PHT32 & 2  & 15 & 60  & 15  & 8  & 31.25 & 4 & 13 & 15 & 3   \\
C\_90 & Chopping & PHT32 & 3  & 3  & 60  & 46  & 8  & 31.25 & 4 & 13 & 15 & 3   \\
\hline
C\_90 & Staring  & PHT22 & 18 & 18 & 69  & 69  & 64 & 31.25 & -- & -- & -- & -- \\ \hline
FCS   & Staring  & PHT32 & -- & -- & --  & --  & 64 & 31.25 & -- & -- & -- & -- \\ \hline
\end{tabular}
\end{center}
\caption{Main parameters for the observations: Filter (see Klaas 
et al. \cite{Kl}), Mode - observing mode, AOT - Astronomical Observing Template 
(see Klaas et al. \cite{Kl}), nY (number of raster points along the
spacecraft Y axis), nZ (number of raster points along the spacecraft Z axis),
dY (step size along the spacecraft Y axis), dZ (step size along the spacecraft 
Z axis), NRPR (number of read-outs per integration ramp), $t_{read-out}$ 
(time per read-out interval), NRCP (number of integration ramps per chopper
plateau), NCHP (number of chopper steps per pointing), CHST (chopper step
size) and NSWP (integer number of chopper sweeps per pointing). The first
four measurement types are the large and small science rasters for both
bands, the next two types are the science rasters for the calibration
measurements, then the measurement type for the Lockman Hole data 
and finally the parameters for the FCS measurements are listed}
\end{table*}

As it will be shown below (Sect. 3), the detected sources have a peak
intensity of only a few percent of the background level. Since the sky in the
survey region is assumed to be very flat, the detectors are illuminated by 
a nearly constant flux for many hours during these observations. As a 
consequence transient and memory effects of the detectors are less 
significant than in other ISO observing modes, where the satellite is 
moving from the background to bright sources observed for a short time not allowing the detectors to stabilize. 

In the following we will present the main instrumental effects which
must be considered when reducing ISOPHOT C100 raster map data. It is important
to understand, that the ISOPHOT data are measurements of the 
voltage on the integrating preamplifiers as a function of time. After a
specified number of non-destructive read-outs (NRPR$- 1$ in Table 2), the 
amplifier voltage is reset with a destructive read-out yielding altogether one 
integration ramp. The change of this voltage with time i.e. the slope of an 
integration ramp is -- idealy -- proportional to the incoming flux.

\subsection{Non-linearity effects}

The Cold Readout Electronics (CRE) of the ISOPHOT instrument exhibit 
non-linearities i.e. the integration ramps deviate systematically
from a linear fit depending on the absolute voltage level of the
individual read-outs in the ramp. This effect has an impact on comparisons 
between integration ramps in different parts of the dynamic range 
(Acosta-Pulido \& Schulz \cite{Ac96}). We have investigated this effect 
in our data by stacking the measurements according to the read-out number 
of each ramp and then determining the median value for each read-out number.

Our data indicate that the deviation from linearity is generally less
than 1\% of the signal. The non-linearity correction implemented in
the IDL based ISO{\bf P}HOT {\bf I}n\-ter\-ac\-ti\-ve {\bf A}nalysis 
software package (PIA) (Gabriel et al. \cite{gab}) is based on a large 
number of measurements covering the entire dynamic range of the amplifiers. 
\begin{figure}[ht]
\begin{center}
\leavevmode
\hbox{%
\epsfysize=60mm
\epsfbox{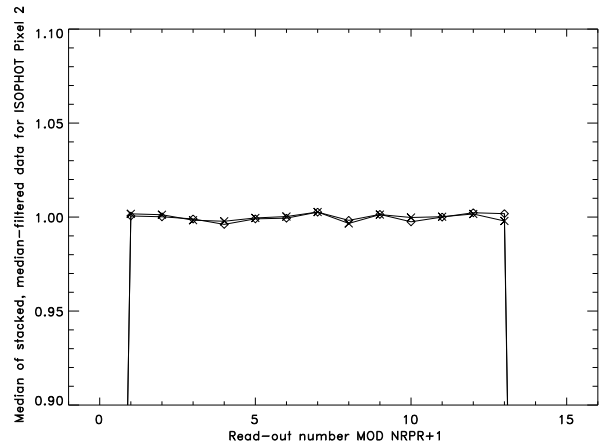}}
\end{center}
\caption{Illustration of the very low level of non-linearity of the read-out electronics for pixel 2 in one of our a C\_60 raster maps (NRPR $= 16$). Here
the first and the two last read-outs per ramp have been discarded as is also 
the case in our data reduction. Crosses indicate data with PIA non-linearity 
correction applied and diamonds without correction. Before stacking the read-outs the data stream has been divided by a median filtered version of 
the data in order to remove drift (see Sect. 2.2) resulting in data values 
around 1}
\end{figure}
As illustrated by a typical example in Fig. 2 we find that within
the part of the dynamical range occupied by our data no effect is
visible.

\subsection{Responsivity drift of the individual C100 pixels}

As emphasized by Lemke et al. (\cite{Le96}) the responsivity of the C100 
pixels can vary in a reproducable way depending on preillumination
history and the accumulated dose of ionizing radiation encountered in
orbit. As shown below (Sect. 3), the peak intensity of the point
sources detected in this survey is only a few per cent of the sky brightness, 
so it is crucial to correct the responsivity drift with good accuracy. 
In Fig. 3 and 4 typical examples of the time ordered data stream for 
C100 pixels are shown on small and large timescales respectively. 
\begin{figure}[ht]
\begin{center}
\leavevmode
\hbox{%
\epsfysize=60mm
\epsfbox{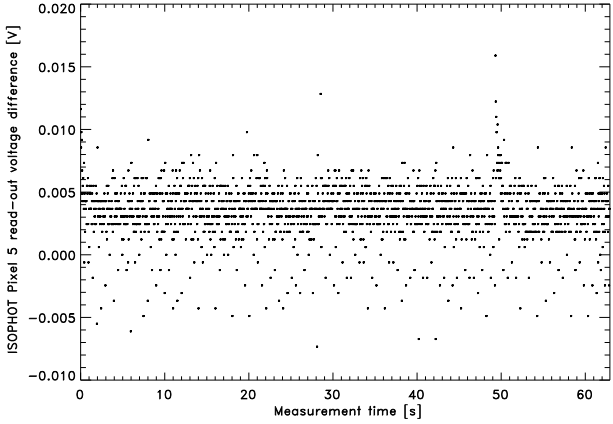}}
\end{center}
\caption{An example of the temporal behaviour of one of the pixels in
the C100 array. The read-out voltage difference between consecutive read-outs
as a function of time exhibits no obvious drift. Note the discrete nature of 
the read-out voltages caused by the finite digital resolution of the read-out
electronics. Note also a glitch event (see Sect. 2.3) at a measurement time 
of $\sim$ 50 s}
\end{figure}

One thing to notice in Fig. 3 is the discrete nature of the read-out
voltages caused by the finite resolution of the read-out electronics. Here the
dynamic range of 2.5 V are digitized using twelve bits or 4096 values resulting
in voltage steps dU = 2.5 V / 4096 = 0.61 mV. In the data reduction
this discretisation is however of no major concern since the signal is
obtained by fitting a first order polynomial to each integration ramp.
In Fig. 3 one further notices a glitch event at a measurement time of 50 s.
These events are discussed further in Sect. 2.3.
\begin{figure}[ht]
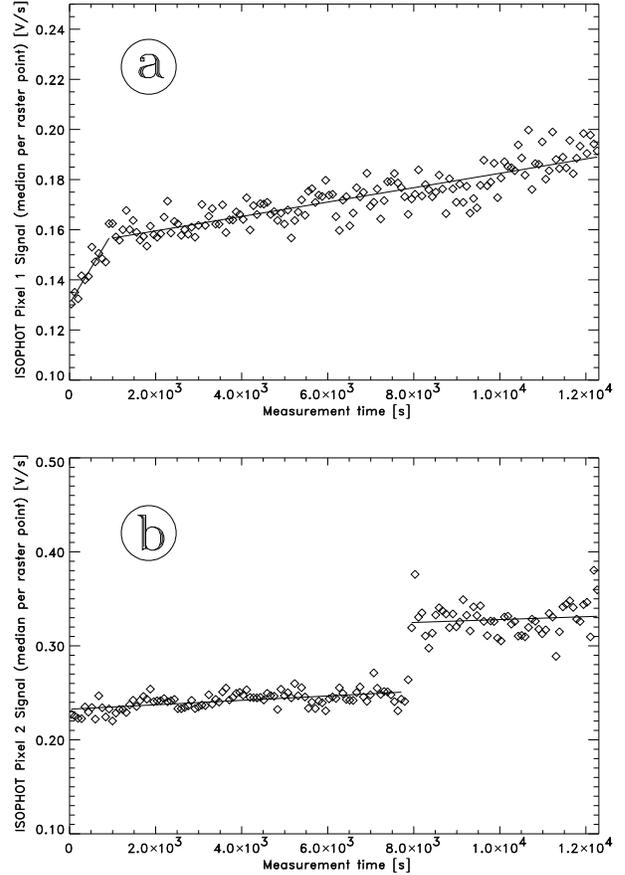

\begin{center}
\leavevmode
\hbox{%
\epsfysize=60mm
\epsfbox{9369.f4a}}
\leavevmode
\hbox{%
\epsfysize=60mm
\epsfbox{9369.f4b}}
\end{center}
\caption{Two examples of the temporal behaviour of the detector signal 
(median value per raster point) of some of the pixels in the 
C100 array. In {\bf a} a typical behaviour with initial, steeply rising
transient, break and slow rise is shown whereas {\bf b} is an example of 
a somewhat more bizarre behaviour with an abrupt jump in the signal level
and increased scatter around the linear fit after the jump}
\end{figure}

It can be seen that variations exist mainly on long
timescales which are comparable with the duration of the entire measurement 
(Fig. 4a). On much shorter timescales, as shown in Fig. 3, no
obvious drift is detectable. 
As it can be seen from Fig. 4a there is also an initial transient period 
where the detector stabilizes. We have performed a detailed
investigation of this behaviour by inspecting the measurement timeline
for each pixel in each raster represented by the median signal per raster 
point as a function of time. In most cases a characteristic behaviour emerges 
as illustrated in Fig. 4a. The signal first exhibits a transient with a steep 
rise, a sharp break and then a more slowly increasing baseline level. Pixels
1 and 5 demonstrate this behaviour all the time whereas other pixels 
-- especially pixel 6 -- are somewhat more unpredictable. The frequency
of rasters with a well-defined break and the related mean time for the
break are listed for each pixel in Table 3. Note that the break time
is defined as the time corresponding to the point where the
two fitted first order polynomials intersect (see Fig. 4a) and the 
frequency is given in percent of the total number of rasters in each 
band (i.e. 13). The distributions of break times and slopes of the first 
order polynomial fitted to the initial transients are shown for pixel 5 
in Fig. 5.
\begin{table}
\caption{Percentage of rasters with {\em baseline-break} and mean
time for break. The last column gives the mean scatter around the long
linear part (see Fig. 4a) for all 26 raster maps. This scatter is
in percent of the start-level for the linear part}
\begin{center}
\begin{tabular}{crrrrc} 
 & \multicolumn{2}{c}{C\_60 data} & \multicolumn{2}{c}{C\_90 data} &
\\ \hline
Pixel & \% & $t_{break}$ [s] & \% & $t_{break}$ [s] & Scatter [\%] \\ \hline  
1 & 100 & 704 & 100 & 352 & 3.3 \\
2 &  54 & 800 &  54 & 358 & 3.3 \\
3 &  69 & 647 &  92 & 365 & 3.4 \\
4 &  77 & 402 &  92 & 248 & 3.0 \\
5 & 100 & 628 & 100 & 284 & 3.8 \\
6 &  23 & 820 &   8 & 322 & 6.3 \\
7 &  77 & 662 &  92 & 260 & 2.9 \\
8 &  62 & 459 & 100 & 251 & 2.8 \\
9 & 100 & 655 &  92 & 369 & 4.0 \\ \hline
\end{tabular}
\end{center}
\end{table}

\begin{figure}[ht]
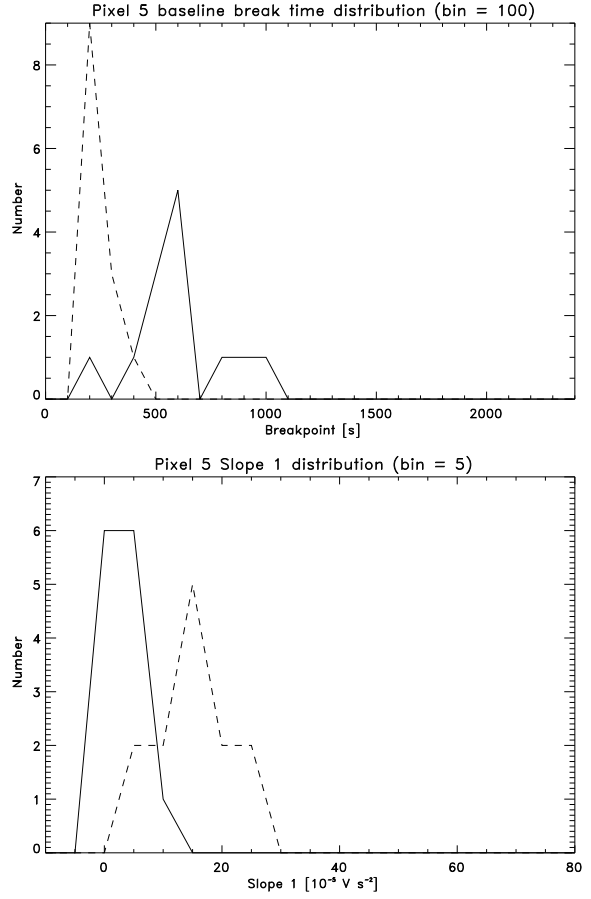

\begin{center}
\leavevmode
\hbox{%
\epsfysize=60mm
\epsfbox{9369.f5a}}
\leavevmode
\hbox{%
\epsfysize=60mm
\epsfbox{9369.f5b}}
\end{center}
\caption{Distributions of baseline break times and slopes of the
1st order polynomium fit to the initial transient. Solid lines represent
C\_60 data and dashed lines C\_90 data. It is obvious that there is
a systematic difference between data obtained in the two bands. This behaviour
may be related to the fact that the C\_60 data have twice as many 
read-outs per integration ramp as the C\_90 data}
\end{figure}

The initial transient period lasts 642 $\pm 138$ s 
for the C\_60 data and 312 $\pm 52$ s for the C\_90 data. This factor of two 
may be related to the fact that the C\_60 data have twice as many read-outs per 
integration ramp as the C\_90 data.

Distortions of the data like the abrupt jump seen in Fig. 4b are not
very frequent. Out of the 26 raster maps such behaviour is seen
four times for pixels 4 and 9, two times for pixels 2, 5, 7, 8, once for
pixel 3 and never for pixels 1 and 6.    

The average increase in signal from the start of the measurement until
the breakpoint is $\sim15$\% of the initial level for the C\_60 data and 
$\sim20$\% for the C\_90 data.
For the remaining part of the measurement the signal increase is $\sim15$\% of
the initial level for the C\_60 data and $\sim20$\% for the C\_90 data.\\

With the responsivity variations seen in Fig. 3 and 4, it is clear, that 
a simple box filtering of the data will not be an appropriate technique.
We have instead employed a median filtering technique in our data reduction to
minimize the influence of large deviations. This method -- described below in
Sect. 3 -- has demonstrated a good ability to remove effects like the ones
shown in Figs. 4a and 4b allowing these data to be included in the
final mapping.

\subsection{Glitches affecting the data stream}

Also glitches induced by cosmic ray events have to be removed from the
data. In Fig. 6 we show a typical example from the time ordered data stream 
of the effect of such a glitch. 
\begin{figure}[ht]
\begin{center}
\leavevmode
\hbox{%
\epsfysize=60mm
\epsfbox{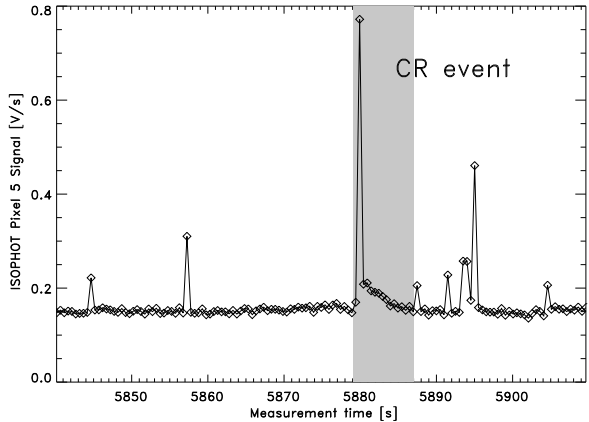}}
\end{center}
\caption{Example of the effect of a cosmic ray induced glitch in the
data stream (signal-level). The grayshaded region represents a subjective
estimate of which signals are affected by the event, but a consistent
method for such a discrimination has not yet been established}
\end{figure}

We have performed a detailed analysis of glitches in our data in order to 
establish important quantities such as the overall glitch rate and the glitch
rate per pixel. The data have been analysed per raster point by determining
mean and $\sigma$ for a gaussian fitted to the distribution of voltage
differences of consecutive read-outs per pointing. All voltage differences 
3$\sigma$ above -- and only above -- the mean are regarded as glitches 
(see Linden-V{\o}rnle \cite{lv98} for details) yielding the
glitch rates listed in Table 4. Here it can be seen that each pixel
is affected about once every 2 s meaning that almost every chopper-plateau
in the C\_60 data and every second in the C\_90 data and the standard
star data is affected at least once by a glitch. It is furthermore worth
noticing that there are significant systematic differences between the glitch 
rates of the individual pixels -- an effect most likely related to 
spontaneous spiking. 
\begin{table}[h]
\caption{Glitch rates for the HUNNSA57 data. First the total rate and the
rate per pixel assuming that each pixel is hit the same number of times 
(i.e. total rate $\times$ 1/9) is given. Then the rates per filter are given 
and finally the rates for the individual C100 pixels are listed. These last
values represent the mean and scatter over all raster maps i.e. both C\_60 
and C\_90 data}
\begin{center}
\begin{tabular}{lc} \hline
All data:            & 6.43 s$^{-1}$ \\
All data per pixel:  & 0.71 s$^{-1}$ \\ \hline \hline
C\_60 filter:        & 6.56 s$^{-1}$ \\
C\_90 filter:        & 6.06 s$^{-1}$ \\ \hline
\end{tabular}
\begin{tabular}{cc} \hline
Pixel    &  rate [s$^{-1}$] \\ \hline
  1      &   $0.66 \pm 0.11$ \\
  2      &   $1.49 \pm 0.19$ \\
  3      &   $0.64 \pm 0.12$ \\
  4      &   $0.95 \pm 0.21$ \\
  5      &   $0.69 \pm 0.10$ \\ 
  6      &   $0.59 \pm 0.08$ \\
  7      &   $0.60 \pm 0.10$ \\
  8      &   $0.23 \pm 0.02$ \\
  9      &   $0.40 \pm 0.08$ \\ \hline
\end{tabular}
\end{center}
\end{table}

It is worth noticing that pixel 8 has a significantly lower
glitch rate. The reason for this is probably that this pixel generally has
a higher scatter in the voltage differences of consecutive read-outs.
This results in a higher threshold for glitch detection and therefore
a lower rate for the weakest -- and most frequent -- glitches.
The lower rate for pixel 8 is therefore not an indication
of this pixel being less noisy -- rather the opposite situation
is the case. 

The behaviour of individual pixels just after a glitch is very unpredictable. 
Although many efforts have been made, we have not yet succeeded
in establishing any usefull correlation e.g. between the read-out voltage
amplitude of the initial jump and the time evolution of the transient 
following immediately after the glitch. Even after a strong glitch the 
read-outs after the event sometimes seem well-behaved compared to the 
read-outs just before the glitch, while in other cases a glitch of similar 
strength, affects many subsequent read-outs.  

As a consequence, no attempts have been made to correct the data 
affected by glitches and instead we have tried to optimize our reduction
in order to assure a minimal influence from the glitches on the
final maps.

\subsection{Chopper dependent signal offset}
We have also used our data to determine the signal offset
of the different pixels at different chopper positions. This effect
is caused by slightly unbalanced beams in both chopper end positions. 
Again we are exploiting the fact that our observations are presumably 
performed under very stable conditions, that there are only a few point 
sources per raster, and they are all faint relative to the sky background. 
\begin{figure}[ht]
\begin{center}
\leavevmode
\hbox{%
\epsfysize=60mm
\epsfbox{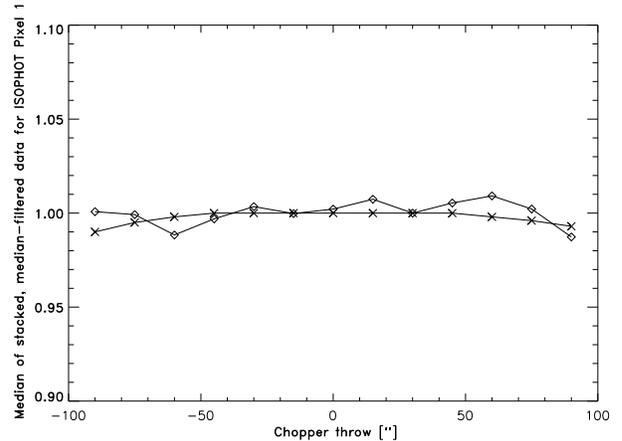}}
\end{center}
\caption{Example of chopper dependent signal offset for pixel 1 in the 
C100 array. Diamonds indicate the offset found in our data whereas crosses 
represent the standard PIA signal offset correction values}
\end{figure}
After the removal of the drift we stack the measurements for each raster
and pixel according to the chopper position and then determine the 
median for each chopper position. In Fig. 7 a typical example can be seen in
comparison with the standard PIA signal offset correction factors. 
In general the effect of the chopper dependent signal offset is in the order 
of 1\% or lower and has therefore not been included in our data reduction 
scheme.

\section{Data reduction}

Following the presentation of the various instrumental effects in the 
previous section we here briefly outline our data reduction scheme. The
backbone of the reduction is PIA version 7.3.1(e) where the processing
is performed on various data-levels, namely: Edited Raw Data (ERD), 
Signal Raw Data (SRD), Signal per Chopper Plateau data (SCP) and
finally Astronomical Analysis Processing (AAP). We have used batch 
processing from ERD- to AAP-level using the processing steps listed in Table 5.
\begin{table}
\caption{Processing steps used in the PIA batch reduction from ERD- to
AAP-level}
\begin{center}
\begin{tabular}{ll} \hline
Level & Processing step\\ \hline
ERD & Two-threshold deglitching \\
ERD$\rightarrow$SRD & 1. order pol. fit to ramps \\
SRD & Reset interval correction \\
SRD & Signal deglitching \\
SRD & Orbital dependent dark current subtraction \\
SRD$\rightarrow$SCP & No drift handling \\
SCP$\rightarrow$AAP & Actual responsivity flux-calibration \\
 & using only FCS1 in the FCS-measurement \\
 & following the science raster \\ \hline
\end{tabular}
\end{center}
\end{table}
Since our data do not seem to be strongly affected by non-linearities 
(see Sect. 2.1) this correction is omitted at the ERD-level. Similarly we 
have not performed signal offset correction on the SCP-level. 
We have furthermore 
reduced the data from SRD- to SCP-level not using the PIA drift handling 
option since we have employed our own non-PIA IDL routine to remove the 
baseline drift. This routine removes the drift in two steps. First the time 
ordered AAP-data are filtered using a narrow median window in 
order to remove remaining data affected by glitches. After this removal of 
high frequency noise we apply a much broader median filter and 
then divide the data with the filtered version in order to remove the large 
scale drift thus flatfielding the data. The map is then rescaled to sky
brightness using the mean value of the sky brightness from each of 
the nine pixels. We have carefully optimized the width of the drift filter 
in order to secure that the statistical properties of the data are generally 
preserved after filtering. 

After drift handling is done for the AAP-level data, PIA is used to produce 
maps. We have chosen to exclude data from pixels 6 and 8 due to the fact
that pixel 6 generaly exhibits very unstable performance and pixel 8 as a rule
is about a factor of two more noisy than the other pixels. For
the mapping we use $15 \times 23$ $\rm arcsec^{2}$ pixels 
which areawise is about 1/6 of a detector pixel. This pixel size is chosen 
since it naturally arises from the combination of 15 arcsec chopper movement
steps and the raster step size along the spacecraft Z axis (2/3 oversampling).
\begin{figure*}[ht]
\begin{center}
\leavevmode
\hbox{%
\epsfysize=100mm
\epsfbox{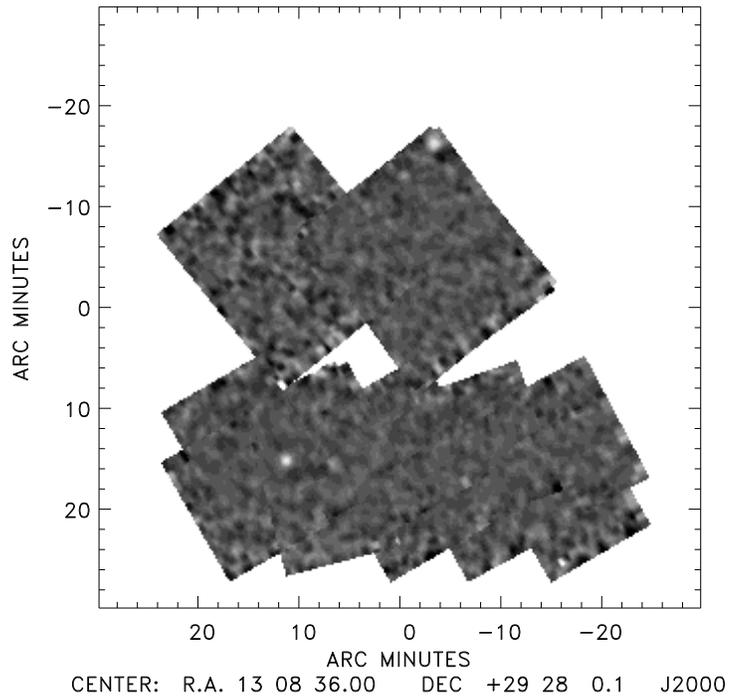}}
\leavevmode
\hbox{%
\epsfysize=100mm
\epsfbox{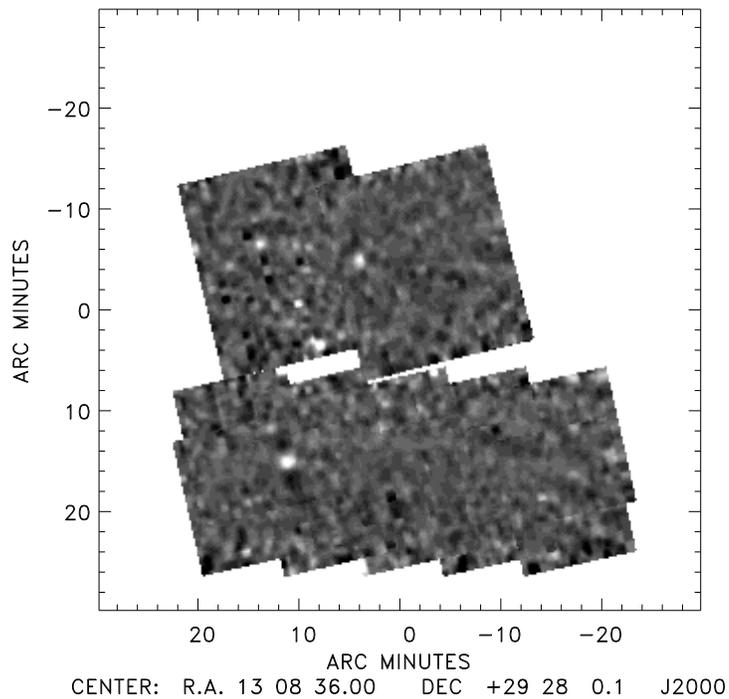}}
\end{center}
\caption{Coadded maps at 60 $\mu$m (top) and 90 $\mu$m (bottom)}
\end{figure*}

\subsection{Source extraction}

In order to detect individual sources we have employed the SExtractor software,
verion 1.2 (Bertin \& Arnouts \cite{ber2}). Each raster map has been 
analysed with the parameters given in Table 6. These parameters have been 
established by investigating the parameter space for the source extraction for 
the repeated large rasters (SA57\_1\_1 in Table 1) which have almost 100\% 
area overlap. The requirement for the parameters is that they should
yield consistent results while minimizing the detection area and $\sigma$-level
constraints i.e. going as faint as possible. 

For a source detection to be classified as reliable we set the requirement 
that in areas where the raster maps are overlapping a detection must be 
confirmed in the overlapping raster map(s) observed in the 
same band within an error-ellipse with radius 10 arcsec. Non-confirmed 
sources in overlapping areas are presently considered to be dubious detections.
In areas without any overlap we must rely on the
robustness of the extraction parameters chosen. Due to the increasing
noise at the edges of the maps detections closer than 45 arcsec to the
edges are however rejected. 

We have coadded the raster maps (see Fig. 8) resulting in a decrease
in noise in the overlapping areas. Running SExtractor on these
coadded maps results in more detections but only in high-noise i.e.
non-overlapping regions. This effect is due to the overall reduction of
the scatter in the background resulting in a lower detection threshold.
\begin{table}
\caption{Parameters used in the source detection using SExtractor version 1.2}
\begin{center}
\begin{tabular}{lr} \hline
Parameter & Value \\ \hline
DETECT\_MINAREA & 9 \\
DETECT\_TRESH & 2.25 \\
BACK\_SIZE & 64 \\ \hline
\end{tabular}
\end{center}
\end{table}

With the chosen SExtractor parameters listed in Table 6 the detected 
sources have a formal signal to noise ratio, SNR $\geq 6.75$. 

\subsection{Flux calibration}

As indicated in Table 5 the flux calibration is done using the internal 
Fine Calibration Sources (FCS). We have analysed the FCS measurements performed 
both before and after the science raster using PIA and the settings listed in 
Table 5 (up to the SCP-level) and we find that the scatter in the derived sky 
brighness is significantly smaller for the FCS measurements performed after the 
science raster. As a result only the FCS measurement following the science 
raster has been used in our batch-processing to derive actual responsivities. 

In order to correct the source signal found within the detection area to obtain
the total flux, we have simulated the PHT32 mapping procedure assuming a linear
response and then used it on the point spread functions for C100 with C\_60 and
C\_90 filters, taken from the calibration file PC1FOOTP.FITS. In Fig. 9 the 
resulting relations between the size of the detection area (number of 
$15 \times 23$ arcsec$^2$ pixels) and the fraction of the total flux for 
both C\_60 and C\_90 is shown. 
\begin{figure}[ht]
\begin{center}
\leavevmode
\hbox{%
\epsfysize=60mm
\epsfbox{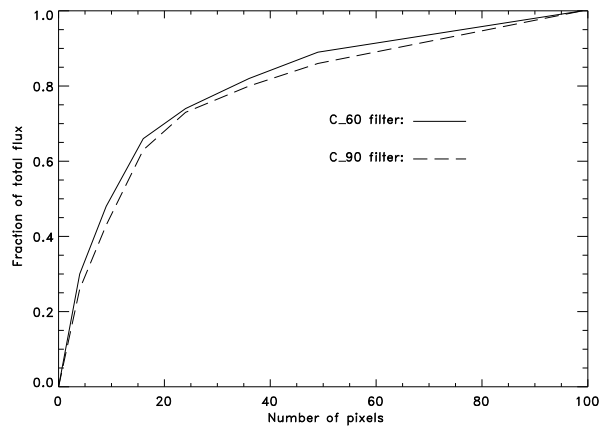}}
\end{center}
\caption{Relation between detection area in terms of $15 \times 23$ arcsec$^2$ 
sky pixels and fraction of total flux for both filters}
\end{figure}

In Table 7 the resulting sky brightnesses are listed and compared to
the COBE DIRBE (Hauser et al. \cite{hau97}) annual average values. We
have verified that the comparison between ISOPHOT and DIRBE
annual average values is viable by checking the DIRBE/ISOPHOT ratio
for DIRBE weekly maps, which include the variable contribution
from the Zodiacal Light. The ratios thus obtained are very similar to 
those listed in Table 7 with a scatter of only 7\% for C\_60 and 3\% for C\_90.
In total it can be seen that the ISOPHOT and DIRBE surface brightness values 
are in agreement within about 10\% for C\_60 and 25\% for C\_90. 
\begin{table}
\caption{Comparison of the COBE DIRBE and the ISOPHOT (PHT) 
sky brightness [MJy sr$^{-1}$]. D/P is the DIRBE/ISOPHOT ratio}
\begin{center}
\begin{tabular}{lrrc} \hline
Sky at target & DIRBE & PHT & D/P \\ \hline
$\alpha$ Aql, C\_60 & 12.85 $\pm$0.34 & 14.36 $\pm$1.32 & 0.90 \\
HR 1654, C\_90      &  4.79 $\pm$0.10 &  6.52 $\pm$0.36 & 0.74 \\
HUNNSA57, C\_60     &  9.91 $\pm$0.10 &  8.78 $\pm$0.64 & 1.12 \\ 
HUNNSA57, C\_90     &  4.72 $\pm$0.04 &  6.15 $\pm$0.20 & 0.77 \\ 
Lockman Hole, C\_90 &  3.93 $\pm$0.08 &  5.22 $\pm$0.20 & 0.75 \\ \hline
\end{tabular}
\end{center}
\end{table}

In order to calibrate the source signals found by SExtractor to an absolute 
flux scale we have used observations of $\alpha$ Aql (C\_60) and HR 1654 
(C\_90), observed with nearly the same instrument parameters as the HUNNSA57 
raster maps (see Table 2). Both stars have been detected by IRAS with a
60 $\mu$m flux for $\alpha$ Aql of 1250 mJy $\pm$11\% (IRAS PSC; flux
quality = 3) and a 100 $\mu$m flux for HR 1654 of 579 mJy $\pm$24\% 
(IRAS FCS; flux quality = 2). More recently a model flux for $\alpha$ Aql 
has however been established as part of the ISO calibration ground-based 
preparatory programme (GBPP, Jourdain de Mui\-zon \& Habing \cite{Jou92} and 
van der Bliek et al. \cite{bliek96}) and similarly a C\_90 model flux for 
HR 1654 has been found by fitting the shape of an appropriate red giant 
branch spectral energy distribution (SED) model to near-IR and mid-IR 
observations obtained for this star, thus extending the SED out to 
300 $\mu$m (Cohen et al. \cite{co96}, Cohen private communication). 
This way the following model fluxes have been obtained: 902.9 mJy and 
712.6 mJy, for $\alpha$ Aql (C\_60) and HR 1654 (C\_90) respectively. We
have chosen to adopt these model fluxes for use in our calibration.

In order to obtain these model fluxes from our observed source signals of the 
standard stars, using the correction for source signal lying outside the 
detection area given in Fig. 9, we find a factor of 2.51 for C\_60 and 1.41 
for C\_90 has to be applied. 

Based on the detection criteria given in Table 6 and the empirical 
calibration factors established above our source detection limit with a 
formal SNR = 6.75 is 90 mJy for C\_60 and 50 mJy for C\_90. 

It is well known that the instant response to a change in flux for 
photoconductors such as the ISOPHOT C100 detector usually is only a
fraction of the stabilized response. In our observations the
detector is observing the sky background for a long time allowing the
detector to reach the stabilized response. When the detector moves onto
a weak point source and then returns to the sky again there has not been
enough on-source time to reach the stabilized response for the
source. It is therefore not unexpected that point source calibration factors 
as those established above would emerge as part of the calibration.

It is however not easily possible to test these empirically established point
source calibration factors of 2.51 for 60 $\mu$m and 1.41 for 90 $\mu$m.
In order to do so one could imagine simulations where sources of known 
flux are introduced in the data and then extracted using the same procedure
as for the real sources. This could in principle verify the validity
of the correction factors and yield a completeness correction as a function
of flux. In order to perform a realistic simlulation like that, systematic
effects at all levels of the data reduction would have to be understood. 
Since the noise properties of the resulting sky maps are 
dominated by highly correlated effects emanating from glitch residuals and
since no detailed understanding of glitch behaviour is currently available, 
we believe that such simlulations are not feasible at the present time.

In Table 8 the reliably detected HUNNSA57 sources with astrometry and 
photometry are listed. For each source $n(60\mu{\rm m})$ and 
$n(90\mu{\rm m})$ indicate the number of detections in independent
rasters over number of possible detections in these rasters (i.e. 2/2 means
two detections out of two possible). For sources with more than one
detection the position and fluxes listed is the mean of the values from
individual detections. For these multiple detections the scatter of the 
individual detections is given as $sc_{\alpha}$ and $sc_{\delta}$ for
the position (in arcsec) and $sc_{f_{\nu}(60\mu{\rm m})}$ and
$sc_{f_{\nu}(90\mu{\rm m})}$ for the fluxes (in mJy). For sources only
detected in one band upper limits are given for the other band.
These limits have been calculated using the noise estimate from
SExtractor for the relevant map and the criteria listed in Table 6.

Two of the sources, HUNNSA57\_1 and HUNNSA57\_6, are IRAS Faint Sources 
with good 60 $\mu$m fluxes (IRAS quality flag 3). The IRAS
fluxes are compared with our values in Table 9 and it can be seen that the
agreement is good within $\sim$ 10\%. 

Based on the results listed in Table 7 and 9 we are confident that we have 
established a reliable calibration which yields a sky brightness in agreement
with COBE DIRBE results and point source fluxes consistent with IRAS FSC at 
least for 60 $\mu$m. IRAS F13059+3000 also has a good 100 $\mu$m flux
but unfortunately this source lies just outside our C\_90 raster maps
prohibiting an independent check of our 90 $\mu$m calibration. Based on
the agreement with DIRBE and the scatter of source fluxes from independent
maps (see Table 8) we estimate that our photometric uncertainty for
point sources is about 25\%. 
\begin{table*}
\caption{Astrometry and photometry (in mJy) for the reliably detected 
sources in the HUNNSA57 and Lockman Hole data. For the HUNNSA57 data 
$n(60\mu{\rm m})$ and $n(90\mu{\rm m})$ gives the number of detections in
independent rasters over number of possible detections (i.e. 2/2 means two 
detections out of two possible). 
For sources with two or more detections in the same band position and flux
is given as the mean of the values from the individual detections. For these
multiple detections the scatter, $sc$, of the individual values is given in 
arcsec for the positions and mJy for the fluxes. For sources only seen in one 
band upper limit for the flux in the other band based on the detection criteria 
listed in Table 6 is given. Note that no 90 $\mu$m flux upper limit is given
for HUNNSA57\_6 since this source lies outside the area covered in this band. 
Since there is no redundancy in the Lockman Hole 
data only SNR values for each source is given ($\sigma = 17$ mJy). One
of the Lockman Hole sources has been detected by IRAS. For this source
the IRAS flux is given}
\begin{center}
\begin{tabular}{lcccccccccc} \hline
Source name & $n(60\mu{\rm m})$ & $n(90\mu{\rm m})$ & $\alpha$(2000) & $sc_{\alpha}$ & $\delta$(2000)& 
$sc_{\delta}$ & $f_{\nu}(60\mu\mbox{m})$ & $sc_{f_{\nu}(60\mu{\rm m})}$
& $f_{\nu}(90\mu\mbox{m})$ & $sc_{f_{\nu}(90\mu{\rm m})}$ \\ \hline
HUNNSA57\_1 & 3/3 & 3/3 & 13 09 28 & $2\arcsec$ & 29 12 47 & $5\arcsec$ &    243  & 61 &   173  & 13 \\
HUNNSA57\_2 & 0/2 & 2/2 & 13 08 55 & $1\arcsec$ & 29 32 54 & $4\arcsec$ & $<  90$ &    &   178  & 55 \\
HUNNSA57\_3 & 0/1 & 1/1 & 13 09 40 &            & 29 34 34 &            & $< 110$ &    &    98  &    \\
HUNNSA57\_4 & 0/1 & 1/1 & 13 08 44 &            & 29 18 50 &            & $<  90$ &    &    68  &    \\
HUNNSA57\_5 & 0/1 & 1/1 & 13 08 45 &            & 29 16 02 &            & $<  90$ &    &    82  &    \\ 
HUNNSA57\_6 & 2/2 & 0/0 & 13 08 20 & $7\arcsec$ & 29 44 26 & $8\arcsec$ &    236  & 26 &    --  &    \\ 
HUNNSA57\_7 & 0/1 & 1/1 & 13 09 14 &            & 29 24 31 &            & $< 110$ &    &   162  &    \\
\hline
\end{tabular}

\begin{tabular}{lccccl} \hline
Source name & $\alpha$(2000) & $\delta$(2000) & $f_{\nu}(90\mu\mbox{m})$ & SNR & IRAS $f_{\nu}(100\mu\mbox{m})$ \\ \hline
Lockman NW1\_1 & 10 32 36.9 & 58 08 53.7 & 331 & 20 &      \\
Lockman NW1\_2 & 10 32 04.4 & 58 08 05.1 & 203 & 12 &      \\
Lockman NW2\_1 & 10 36 04.6 & 57 47 57.2 & 265 & 16 &      \\
Lockman E1\_1  & 10 50 51.7 & 57 35 17.0 & 197 & 12 &      \\
Lockman E1\_2  & 10 49 48.1 & 57 34 58.1 & 165 & 10 &      \\ 
Lockman E4\_1  & 10 53 49.8 & 57 07 20.5 & 747 & 45 & 1218 \\ 
Lockman E4\_2  & 10 53 02.7 & 57 05 51.1 & 177 & 15 &      \\
Lockman E4\_3  & 10 52 53.0 & 57 08 16.3 & 203 & 12 &      \\
\hline
\end{tabular}
\end{center}
\end{table*}

\begin{table}
\caption{Flux comparison for IRAS faint sources in the HUNNSA57 sample.
The flux is given in mJy}
\begin{center}
\begin{tabular}{l|cc} \hline
Source name                   & HUNNSA57\_1 & HUNNSA57\_6 \\
IRAS name                     & F13071+2928 & F13059+3000 \\
PHT $f_{\nu}(60\mu\mbox{m})$  & 243 $\pm$61 & 236 $\pm$26 \\
IRAS $f_{\nu}(60\mu\mbox{m})$ & 266 $\pm$42 & 269 $\pm$43 \\ 
PHT/IRAS                      & 0.91        & 0.88        \\ \hline
\end{tabular}
\end{center}
\end{table}

\subsection{Analysis of the ISOPHOT Lockman Hole data}

As mentioned in Sect. 1 we have included the ISOPHOT observations of
a 1.1 deg$^2$ blank field in the Lockman Hole area (Kawara et al. \cite{Ka98})
in this work. We have reduced the C\_90 data using the same batch processing
steps as outlined in Table 5 and also removed drift at the AAP-level
using median filtering. We have adobted a similar median filter width
as Kawara et al. (\cite{Ka98}) namely the number of pointings along a
raster leg $\times$ 1.5. 
Source extraction is also performed using SExtractor with parameters 
similar to those listed in Table 6. Only the parameter for minimum 
detection area, DETECT\_MINAREA, was trimmed to 11 instead of 9. Since there
is no significant overlap between the eight rasters we have not been
able to optimize the extraction parameters in the same way as for the
HUNNSA57 data but rely instead on the robustness of the method.

Regarding calibration we find that the sky brightness in the C\_90 data
resulting from the ISOPHOT FCS calibration yields a sky background which 
is about 30\% bright\-er than the DIRBE annual average value (see Table 7). 
This is in fairly good agreement with our findings for the HUNNSA57 data. 
Regarding point sources, we have simulated the PHT22 mapping mode again
assuming a linear response in order to establish the relation between 
the observed flux and the detection area (number of pixels -- here 
$23 \times 23$ arcsec$^2$). After using this relation to correct the 
source signals found by SExtractor to obtain the total flux we have applied 
our empirical C\_90 calibration factor of 1.41 obtained from the standard 
star observations (see Sect. 3.2). Astrometry and photometry for the 
detected sources is given in the second part of Table 8. Since there is no
redundancy in the Lockman Hole data in the sense of positions and fluxes 
from independent raster maps we only list SNR values for the detected 
sources. As indicated in Table 8, the source Lockman E4\_1 has also been 
detected by IRAS (F10507+5723) with an IRAS 100 $\mu$m flux of 1218 mJy. 
With the calibration established in this work this flux is underestimated 
by 63\%. Kawara et al. (\cite{Ka98}) use this IRAS source to
calibrate their point source fluxes and therefore fluxes and flux limits
presented by Kawara et al. (\cite{Ka98}) have to be corrected by a factor
of $(1.0 + 0.63)^{-1} = 0.61$ in order to allow comparison with our results.

\section{Discussion}

As it can be seen from the coadded HUNNSA57 maps in Fig. 8 point sources are 
sitting on top of a largely structureless background. This is expected since
our median filtering technique suppresses extended structure in the background.
The brightness variations in the individual C\_90 maps have an rms of about 
0.5 MJy sr$^{-1}$ equivalent to 6 mJy per $15 \times 23$ arcsec$^2$ pixel. 
This compares well to the results obtained by Herbstmeier et al. (\cite{he98})
where observations of a smooth area on the sky, M03, selected for studies
of structure in the zodiacal light yields fluctuations at the 6 mJy level
per $46 \times 46$ arcsec$^2$ pixel. Both our value and the value found
by Herbstmeier et al. (\cite{he98}) is considerably higher than the
expected one sigma confusion noise for infrared cirrus of $\sigma_{cirrus}$(90$\mu$m) $\sim$ 0.14 mJy for a $23 \times 23$ arcsec$^2$ pixel presented by Kawara et al. (\cite{Ka98}).  

As it can be seen from Table 8 we have in our HUNNSA57 survey detected seven
sources at 90 $\mu$m brighter than 68 mJy and two sources at 60 $\mu$m brighter 
than 236 mJy. Only one of the sources, HUNNSA57\_1, is detected in both 
bands yielding a flux ratio, 
$\frac{f_{\nu}(90\mu{\rm m})}{f_{\nu}(60\mu{\rm m})} = 0.7$.
For the HH87 60 $\mu$m sources, which also have a moderate-quality 
100 $\mu$m flux, the mean flux ratio $\frac{f_{\nu}(100\mu{\rm m})}{f_{\nu}(60\mu{\rm m})}$ is 3.1 whereas only 8\% of the sources have a 
flux ratio $\sim 1$. One of these sources is the nearby galaxy NGC 6552, 
which is identified as a Seyfert 2 galaxy (see e.g. Bassani 
et al. \cite{bas99}). For a low redshift galaxy, assuming a grain 
emissivity index of 1.5, a flux ratio $\frac{f_{\nu}(90\mu{\rm m})}{f_{\nu}(60\mu{\rm m})} = 0.7$ actually corresponds to a dust 
temperature of $\sim 65$ K whereas a ratio of 3.1 corresponds to a 
temperature of about 25 K. This seems to indicate that our source, 
HUNNSA57\_1, might have a dust component with a temperature of about 65 K, 
maybe harbouring an active galactic nucleus. The detailed characteristics 
of this and the other sources will be discussed further in paper II dealing 
with the individual sources supplemented by optical images and spectra. 

For the observations of the Lockman Hole the number of reliably detected 
sources amounts to eight down to 165 mJy which is somewhat more conservative 
than the result of 36 sources down to 90 mJy reported by Kawara et al. 
(\cite{Ka98}). Based on our experience with the redundant HUNNSA57 data we 
are however -- at this stage -- not confident about detections below our limit
of 165 mJy. Furthermore Kawara et al. (\cite{Ka98}) do not in detail describe 
their criteria for finding sources. Note that the flux limit reported Kawara et al. (\cite{Ka98}) is 150 mJy which here has been corrected by a factor of 0.61 
in order to match our calibration as described in Sect. 3.3.

As mentioned in Sect. 2 the total area is 0.40 deg$^2$ in both bands for
the HUNNSA57 survey. This also takes into account the area lost by rejecting
sources at the edges. For the Lockman Hole data the area is 1.1 deg$^2$ . The 
cumulative number counts at 90 $\mu$m down to 150 mJy are 14.8 per deg$^2$ 
and at 60 $\mu$m 5 per deg$^2$ down to 90 mJy. Compared to the number counts coming from the Guiderdoni et al. (\cite{guid1}) no-evolution model  
i.e. no evolution of neither source density nor luminosity and a cosmology 
with $h = 0.5$ and $\Omega_0 = 1$ our number counts indicate a number
density which is higher by a factor of $1.8\pm0.6$ for 90 $\mu$m but are inconclusive for 60 $\mu$m.
\begin{figure*}[ht]
\begin{center}
\leavevmode
\hbox{%
\epsfysize=110mm
\epsfbox{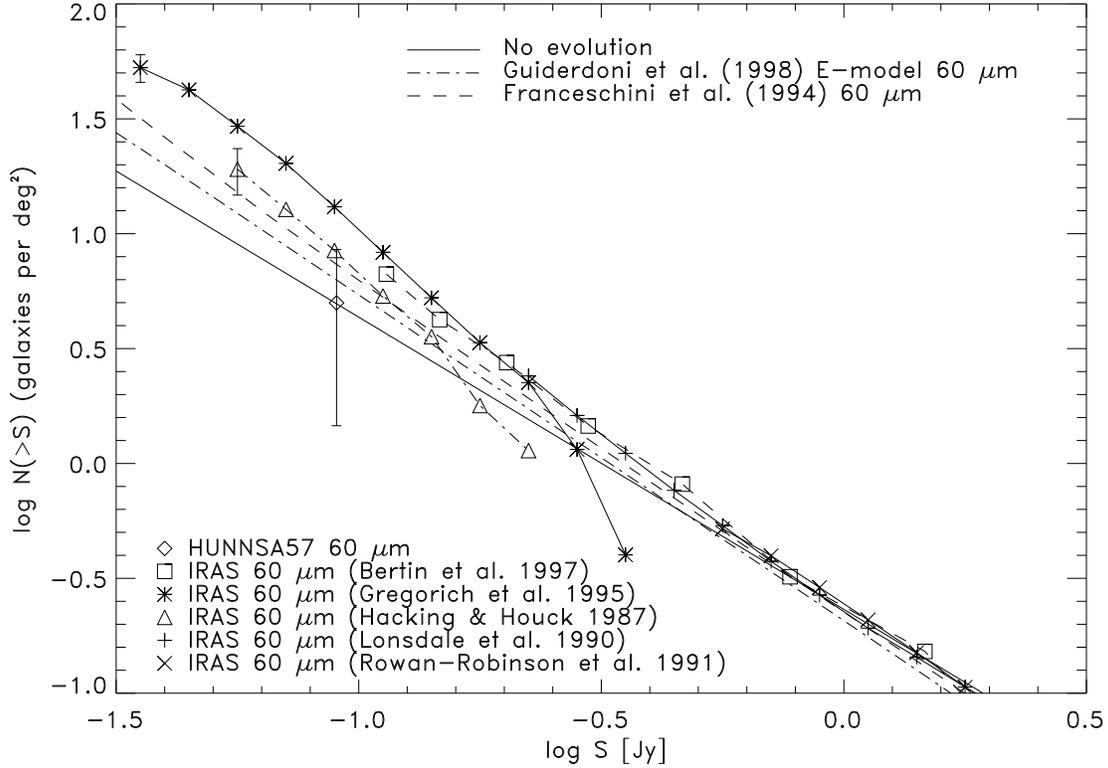}}
\end{center}
\caption{Cumulative 60 $\mu$m number counts for HUNNSA57 along with
previously published source counts and models. Our data are plotted with
with error bars representing Poissonian uncertainties whereas the 
previously published source counts are plotted without error-bars}
\end{figure*}
\begin{figure*}[ht]
\begin{center}
\leavevmode
\hbox{%
\epsfysize=110mm
\epsfbox{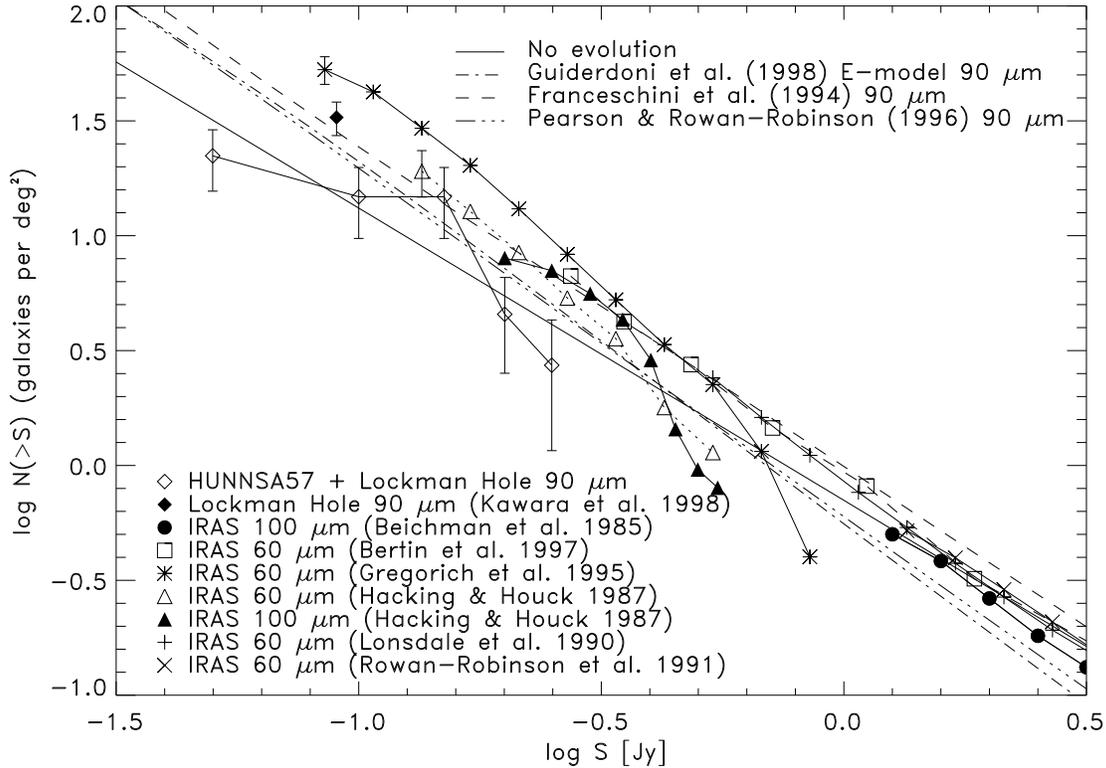}}
\end{center}
\caption{Cumulative 90 $\mu$m number counts for HUNNSA57 + Lockman Hole
data along with previously published counts and models. Except for the
IRAS 100 $\mu$m data from Beichman et al. (\protect{\cite{iras}}) and HH87 the 
previously published counts are actually the 60 $\mu$m counts presented 
in Fig. 10 with the flux scale multiplied by a factor of 2.4. This factor 
has been established as the flux factor between the IRAS 60 $\mu$m and 
100 $\mu$m point source counts at $|b| > 50^{\circ}$ and we have verified 
that this scaling is valid at least for low redshift galaxies ($z \leq 0.5$) 
with dust temperatures in the range 30-40 K and an emissivity indices ranging 
from 1 to 2. As in Fig. 10 our data are plotted with error bars representing 
Poissonian uncertainties whereas the previously published source counts are 
plotted without error bars except for the Kawara et al. (\protect{\cite{Ka98}}) data and the faintest bin of the Hacking \& Houck (\protect{\cite{hack1}}) 
and Gregorich et al. (\protect{\cite{greg1}}) data. Note that the flux limit 
of the Kawara et al. (\protect{\cite{Ka98}}) data has been corrected as discussed in Sect. 4 in order to match up with our calibration}
\end{figure*}

In Fig. 10 the number density of 60 $\mu$m sources from our HUNNSA57 
survey is plotted together with previously published counts from 
Bertin et al. (\cite{ber1}), Gregorich et al. (\cite{greg1}), 
Hacking \& Houck (\cite{hack1}), Lonsdale et al. (\cite{lons}) 
and Rowan-Robinson et al. (\cite{row}). Due to the large number 
of data points the previously published data are plotted without 
error bars. The evolution model from Franceschini et al. (\cite{fran}) 
which includes an early population of dust shrouded galaxies
and the Guiderdoni et al. (\cite{guid1}) E-model which introduces
a rapidly increasing fraction of ULIRGs with redshift are included.
Also the no-evolution model from Guiderdoni et al. (\cite{guid1}) 
mentioned above is included.

In Fig 11 the 90 $\mu$m number densities found from the combined 
HUNNSA57 and Lockman Hole surveys are plotted in a similar way also
including evolutionary models from the references listed above plus 
the model from Pearson \& Rowan-Robinson (\cite{PeRo}) which includes 
a strongly evolving population of dusty starburst galaxies which 
absorb 90-95\% of the optical light and re-emit it at infrared
wavelengths.

In order to compare with source count data we have done three things. 
First we have included 100 $\mu$m counts from the IRAS PSC 
(Beichman et al. \cite{iras}) which are the only published 100 $\mu$m 
number counts from IRAS and secondly we have included source counts of the
HH87 sources with moderate-quality 100 $\mu$m flux excluding of course
NGC 6543 and another source noted by HH87 to be cirrus. It is important
to note that the 100 $\mu$m sources from HH87 do not represent an independent
sample but are instead 100 $\mu$m fluxes determined at the positions
of the detected 60 $\mu$m sources. Finally we have 
multiplied the flux scale of the 60 $\mu$m counts shown in Fig. 10 by a 
factor of 2.4. This factor has been established as the flux factor between 
the IRAS 100 $\mu$m and 60 $\mu$m Point Source counts at $|b| > 50^{\circ}$ 
(Beichman et al. \cite{iras}) at equal number densities. It has
been confirmed to us that such a scaling is viable since the area 
coverage for the 60 $\mu$m and 100 $\mu$m number counts is identical 
(T. Chester, IPAC, private communication). 
The IRAS 100 $\mu$m and ISOPHOT C\_90 filter exhibit fairly similar 
spectral response and we have verified that the scaling factor 2.4 is 
valid to within a few percent at least for low redshift galaxies 
($z \leq 0.5$) with dust temperatures in the range 30-40 K and grain 
emissivity indices ranging from 1 to 2. This also implies that the IRAS 
100 $\mu$m number counts can be directly compared with the ISOPHOT 
90 $\mu$m data. As in Fig. 10 the previously published source count data 
plotted in Fig. 11 are shown without error bars except for the faintest
bin of the Hacking \& Houck (\cite{hack1}) and Gregorich et al. (\cite{greg1})
data.

From Fig. 10 it can be seen that the sparseness of our data do not allow us
to make any strong statements about the 60 $\mu$m population. With our
60 $\mu$m detection limit of 90 mJy at the formal SNR of 6.75 it is
noteworthy that we do not find any sources in the flux range from 236 mJy to
90 mJy. This situation is probably a result of the small survey area
by chance missing out fainter sources.

For the 90 $\mu$m data in Fig. 11 the combined HUNNSA57 and Lockman Hole number 
densities are in line with evolutionary models by Guiderdoni et al. 
(\cite{guid1}), Pearson \& Rowan-Robinson (\cite{PeRo}) and Franceschini et al. 
(\cite{fran}) without being able to discriminate between them. Also the
no-evolution scenario is not ruled out by the data even though there is
an indication of evolution at least at 150 mJy. If this single data point is higher just by chance, no-evolution would be the most obvious match to the 
data. 
 
When comparing our 90 $\mu$m source counts with the HH87 100 $\mu$m source counts and the empirically scaled 60 $\mu$m counts there seems to be a fairly good agreement between our results and the results obtained by HH87, Bertin 
et al. (\cite{ber1}), and also the number densities found by Gregorich et 
al. (\cite{greg1}). It is seen that the curve for our cumulated source counts 
is flattening below a flux of 150 mJy. This effect might be a result of 
incompleteness below this level but it could also indicate that the number 
density of sources below this level is actually leveling out. At the brighter
end the number density falls of somewhat more steeply than Bertin et al. 
(\cite{ber1}) which probably is due to the low probability of finding sources
this bright in an area of only a few deg$^2$. This trend is also seen in the 
HH87 100 $\mu$m number counts. 

Puget et al. (\cite{pug1}) and Dole et al. (\cite{dol1}) present
results from the FIRBACK survey at 175 $\mu$m using the ISOPHOT C200 detector.
Puget et al. (\cite{pug1}) discuss the FIR properties of the sources detected
in the survey by performing a crude estimate of the number of sources one would 
expect if the galaxies seen at 175 $\mu$m are induced by the low redshift galaxies 
detected by IRAS at 60 $\mu$m. This estimate is based on the IRAS 60 $\mu$m galaxy
counts as given by Lonsdale et al. (\cite{lons}) and Bertin et al. (\cite{ber1})
and on a constraint on the color 
$\frac{f_{\nu}(175\mu{\rm m})}{f_{\nu}(60\mu{\rm m})} \leq 1.5$ resulting from 
IRAS, ISO and ground based measurements of a few galaxies classified as starbursts or mergers.  
Based on the counts for the entire survey as given by Dole et al. (\cite{dol1}) 
one finds that the extrapolated IRAS counts underpredict the 175 $\mu$m counts by a factor of 5.5. As noted by 
Puget et al. (\cite{pug1}) the ISOPHOT Serendipity Survey which also has 
gathered data at 175 $\mu$m (Stickel et al. \cite{stick}) tends to yield a  
$\frac{f_{\nu}(175\mu{\rm m})}{f_{\nu}(60\mu{\rm m})} \sim 2$. Using this
color ratio the extrapolated 60 $\mu$m IRAS counts end up being a factor $\sim2.5$
lower than the 175 $\mu$m counts. If the number densities found by Gregorich et
al. (\cite{greg1}) are used, still employing
$\frac{f_{\nu}(175\mu{\rm m})}{f_{\nu}(60\mu{\rm m})} = 2$, the factor is even
further reduced to a factor of about 2. This in turn means
that even though we expect the sources found at 175 $\mu$m to have a
higher median redshift than the sources detected by IRAS there might not be
such a large discrepancy in the populations detected as concluded
by Puget et al. (\cite{pug1}). This view is further strengthened by very
recent observations of FIRBACK sources using SCUBA reported by Scott et al. 
(\cite{sc99}). Out of 10 bright 175 $\mu$m sources 4 objects are convincingly
detected at 850 $\mu$m and fits to simple spectral energy distributions
suggest a range of low to moderate redshifts, $z < 1$. It should also 
be mentioned that the counts presented by Dole et al. (\cite{dol1}) are well 
in line with the E-model by Guiderdoni et al. (\cite{guid1}) whereas the 
model by Franceschini et al. (\cite{fran2}) overpredict the counts at the 
bright end.

The recent results from submillimeter surveys at 850 $\mu$m presented
by Blain et al. (\cite{bl}), Hughes et al. (\cite{hu}), Barger et al. (\cite{ba})
and Eales et al. (\cite{ea}) indicate a number density at a flux level of 
about 1 mJy which is more than two orders of magnitude higher than would be 
expected on the basis of a non-evolving local IRAS luminosity function (Smail et al.
\cite{sm98}). The E-model by Guiderdoni et al. (\cite{guid1}) is well in 
line with the number densities found by Blain et al. (\cite{bl}), 
Hughes et al. (\cite{hu}) and Eales et al. (\cite{ea}) whereas the counts 
presented by Barger et al. (\cite{ba}) are a few times too low compared to this
model. The model presented by Franceschini et al. (\cite{fran2}) underpredicts the
counts by a factor of a few. As discussed by Puget et al. (\cite{pug1}) the
source counts at 850 $\mu$m are more sensitive to the details of evolutionary
scenarios, like those discussed by Guiderdoni et al. (\cite{guid1}), than
is the case for the counts at 60 $\mu$m. The reason for this behaviour is threefold:
(1) The rest-frame wavelength range around 100 $\mu$m holds the major part of 
infrared emission due to starformation. (2) The star formation rate increases
strongly between $z = 0$ and $z = 1.5$ and a similar behaviour in the infrared is
likely. (3) The {\it negative} $k$-correction at submillimeter wavelengths favours
the detection of high-$z$ objects. Given this the 850 $\mu$m counts
seem at present to favour the Guiderdoni et al. (\cite{guid1}) E-model
including a rapidly increasing fraction of ULIRGs with redshift.

\section{Conclusion}

Using the ISOPHOT C100 detector we have conducted a deep survey at
60 $\mu$m and 90 $\mu$m in a small part of Selected Area 57 near
the North Galactic Pole. We have produced maps with spatial resolution
much higher than IRAS and have established a calibration which
is in line with both the COBE DIRBE annual average sky background and 
IRAS point sources to within about 25\% at least at 60 $\mu$m.  

Source counts based on the combined source detections both in the HUNNSA57 data
and our reduction of the Lockman Hole 90 $\mu$m survey indicate that evolution 
as given by the models of Guiderdoni et al. (\cite{guid1}) (E-model) and 
Franceschini et al. (\cite{fran}) is consistent with our data. Our data do 
however not permit us to discriminate between these models and do
furthermore not rule out no-evolution scenarios even though there is an 
indication of evolution at least at 150 mJy. Based on the
source counts at 850 $\mu$m presented by e.g. Blain et al. (\cite{bl}) it
seems however that the E-model by Guiderdoni et al. (\cite{guid1}) is 
at present the model best in line with observational evidence at both
FIR and submillimetre wavelengths.   
 
Our 90 $\mu$m source counts are in line with HH87 100 $\mu$m source counts and
the empirically scaled 60 $\mu$m results from  Bertin et al. (\cite{ber1}), HH87
not precluding the number densities found by Gregorich et al. (\cite{greg1}).
  
In paper II a detailed discussion of the individual sour\-ces detected in this
survey will be given focusing on optical identification of the sources by 
means of optical imaging and spectroscopy. 

\begin{acknowledgements} 
The ISOPHOT data presented in this paper were partly reduced using PIA 
Version 7.3.1(e), which is a joint development by the ESA Astrophysics 
Division and the ISOPHOT Consortium.

We would like to thank Bruno Guiderdoni (IAP, France) and Alberto Franceschini
(University of Padova, Italy) for providing us their evolutionary models.

Finally we would like to thank the PHOT Instrument Dedicated Team (ESA/VILSPA,
Madrid) --  especially Bernhard Schulz and Carlos Gabriel -- for their help and 
support.
 
\end{acknowledgements}


\begin{thebibliography}{}
\bibitem[1996]{Ac96}
Acosta-Pulido J.A., Schulz B., 1996, CRE non-linearity correction 
from PV phase calibration report, June 1996.
\bibitem[1996]{ash1}
Ashby M.L.N., Hacking P.B., Houck J.R., Soifer B.T., Weisstein E.W., 1996,
ApJ 456, 428
\bibitem[1999]{ba}
Barger A.J., Cowie L.L., Sanders D.B., 1999, ApJ 518, L5
\bibitem[1999]{bas99}
Bassani L., Dadina M., Maiolino R., et al., 1999, ApJS 121, 473
\bibitem[2000]{bl}
Blain A.W., Ivison R.J., Kneib J.-P., Smail I., 2000, Galaxy counts at 
450 $\mu$m and 850 $\mu$m. In: Bunker A.J., van Breugel W.J.M. (eds.) 
The Hy-redshift universe: Galaxy formation and evolution at high hedshift,
ASP conference series,  vol. 193
\bibitem[1985]{iras}
Beichman C.A., Neugebauer G., Habing H.J., Clegg P.E., Chester T.J. (eds.),
1985, Infrared Astronomical Satellite (IRAS), Catalogs and Atlases,
Explanatory Supplement, GPO, Washington DC
\bibitem[1996]{ber2}
Bertin E., Arnouts S., 1996, A\&AS 117, 393
\bibitem[1997]{ber1}
Bertin E., Dennefeld M., Moshir M., 1997, A\&A 323, 685
\bibitem[1996]{co96}
Cohen M., Witteborn F.C., Carbon D.F., Davies J.K., Wooden D.H., 
Bregman J.D., 1996, AJ 112, 2274
\bibitem[1999]{dol1}
Dole H., Lagache G., Puget J-L., et al., 1999, FIRBACK far infrared
survey with ISO: Data reduction, analysis and first results. In: 
Cox P., Kessler M.F. (eds.) The Universe as seen by ISO, Volume 2, SP-427, 
ESA Publications Division, ESA/ESTEC
\bibitem[1999]{ea}
Eales S., Lilly S., Gear W., et al., 1999, ApJ 515, 518
\bibitem[1994]{fran}
Franceschini A., Mazzei P., De Zotti G., 1994, ApJ 427, 140
\bibitem[1998]{fran2}
Franceschini A., Andreani P., Danese, L., 1998, MNRAS 296, 709
\bibitem[1997]{gab}
Gabriel C., Acosta-Pulido J., Heinrichsen I., Skaley D., Morris H.,
Tai W.-M., 1997, The ISOPHOT Interactive Analysis PIA, a calibration and
scientific analysis tool. In: Hunt G., Payne H.E. (eds.) Proc. of the ADASS VI
conference, ASP conference series,  vol. 125
\bibitem[1995]{greg1}
Gregorich D.T., Neugebauer G., Soifer B.T., Gunn J.E., Herter T.L., 1995,
AJ 110, 259
\bibitem[1998]{guid1}
Guiderdoni B., Hivon E., Bouchet F.R., Maffei B., 1998, MNRAS 295, 877
\bibitem[1987]{hack1}
Hacking P., Houck J.R., 1987, ApJS 63, 311
\bibitem[1987]{hack2}
Hacking P., Condon J.J., Houck J.R., 1987, ApJ 316, L15
\bibitem[1997]{hau97}
Hauser M.G., Kelsall T., Leisawitz D., Weiland J. (eds.), 1997,
COBE Diffuse Infrared  Background Experiment (DIRBE) Explanatory
Supplement, version 2.1, COBE Ref. Pub. No. 97-A, Greenbelt, MD: NASA/GSFC
\bibitem[1998]{he98}
Herbstmeier U., \'{A}brah\'{a}m P., Lemke D., et al., 1998, A\&A 332, 739 
\bibitem[1998]{hu}
Hughes D.H., Serjeant S., Dunlop J., et al., 1998, Nature 394, 241
\bibitem[1992]{Jou92}
Jourdain de Muizon M., Habing H.J., 1992, The ISO ground-based preparatory
programme working group (ISO-GBPPWG). In: Encrenaz Th., Kessler M.F. (eds.) 
Infrared astronomy with ISO, Les Houches series, Nova science publishers Inc.
\bibitem[1998]{Ka98}
Kawara K., Sato Y., Matsuhara H., et al., 1998, A\&A 336, L9
\bibitem[1996]{Ke96}
Kessler M.F., Steinz J.A., Anderegg M.E., et al., 1996, A\&A 315, L27
\bibitem[1994]{Kl}
Klaas U., Kr\"uger H., Heinrichsen I., Heske A., Laureijs R. (eds.), 1994,
ISOPHOT Observer's Manual, version 3.1
\bibitem[1986]{Koo:Kr}
Koo D.C., Kron R.G., Cudworth K.M., 1986, PASP 98, 285
\bibitem[1993]{le93}
Lemke D., Wolf J., Schubert J., Patrashin M., 1993, SPIE 1946, 261
\bibitem[1994]{le94}
Lemke D., Garzon F., Gem\"und H.P., et al., 1994, Opt. Eng. 33(1), 20
\bibitem[1996]{Le96}
Lemke D., Klaas U., Abolins J., et al., 1996, A\&A 315, L64
\bibitem[1998]{lv98}
Linden-V{\o}rnle M.J.D., 1998, Glitch events in deep ISOPHOT C100
observations: A case study, report
\bibitem[1990]{lons}
Lonsdale C.J., Hacking P.B., Conrow T.P., Rowan-Robinson M., 1990, ApJ 358, 60
\bibitem[1999]{man1}
Mann R.G., 1999, Submillimetre source counts: First results from SCUBA. In:
de Oliveira-Costa A., Tegmark M. (eds.) Microwave foregrounds, ASP conference
series, vol. 181
\bibitem[1992]{mosh1}
Moshir M., et al., 1992, Explanatory Supplement to the IRAS Faint Source 
Survey, Version 2, JPL D-10015 8/92
\bibitem[1996]{PeRo}
Pearson C., Rowan-Robinsin M., 1996, MNRAS 283, 174
\bibitem[1999]{pug1}
Puget J-L., Lagache G., Clements D.L., et al., 1999, A\&A 345, 29
\bibitem[1991]{row}
Rowan-Robinson M., Saunders W., Lawrence A., Leech K., 1991, MNRAS 253, 485
\bibitem[1999]{sc99}
Scott D., Lagache G., Borys C., et al., 1999, astro-ph/9910428
\bibitem[1999]{sm98}
Smail I., Ivison R., Blain A., Kneib J.-P., 1999, Deep sub-mm surveys
with SCUBA. In: Holt S., Smith E. (eds.) After the dark ages: When galaxies
were young (the Universe at $2 <$ z $< 5$), American Institute of Physics
Press
\bibitem[1987]{soif1}
Soifer B.T., Houck J.R., Neugebauer G., 1987, ARA\&A 25, 187
\bibitem[1950]{st50}
Stebbins J., Whitford A.E., Johnson H.L., 1950, ApJ 112, 469
\bibitem[1998]{stick}
Stickel M., Bogun S., Lemke D., et al., 1998, A\&A 336, 116 
\bibitem[1996]{bliek96}
van der Bliek N.S., Bouchet P., Habing H.J., et al., 1996, The Messenger 70,
28

\end{thebibliography}
\end{document}